\documentclass[pre,twocolumn,showpacs,preprintnumbers,amsmath,amssymb]{revtex4}
\usepackage{graphics,bm}

\begin{document}

\title{Alternative sampling for variational quantum Monte Carlo}

\author{J. R. Trail}
\email{jrt32@cam.ac.uk}
\affiliation{TCM Group, Cavendish Laboratory, University of Cambridge,
JJ Thomson Avenue, Cambridge, CB3 0HE, UK}

\date{May, 2007}

\begin{abstract}
Expectation values of physical quantities may accurately be obtained by the 
evaluation of integrals within Many-Body Quantum mechanics, and these 
multi-dimensional integrals may be estimated using Monte Carlo methods.
In a previous publication it has been shown that for the simplest, most 
commonly applied strategy in continuum Quantum Monte Carlo, the random error in 
the resulting estimates is not well controlled.
At best the Central Limit theorem is valid in its weakest form, and at worst it 
is invalid and replaced by an alternative Generalised Central Limit theorem and 
non-Normal random error.
In both cases the random error is not controlled.
Here we consider a new `residual sampling strategy' that reintroduces the 
Central Limit Theorem in its strongest form, and provides full control of the 
random error in estimates.
Estimates of the total energy and the variance of the local energy within 
Variational Monte Carlo are considered in detail, and the approach presented 
may be generalised to expectation values of other operators, and to other 
variants of the Quantum Monte Carlo method.
\end{abstract}



\maketitle

A primary problem in solving for the ground states of many body quantum systems 
is the evaluation of $3N$-dimensional integrals, where $N$ is the number of 
particles interacting in $3$-dimensional space.
This paper considers estimates of expectation values of a many body-trial 
wavefunction and operator combinations, with particular emphasis on those used 
for the optimisation of a trial wavefunction via a parameterised freedom within 
that wavefunction.
Monte Carlo (MC) methods provide a powerful numerical tool for evaluating these 
integrals by expressing the exact integral as an expectation value.
By constructing a sample estimate of this expectation value, such problems can 
be made tractable.

The resulting estimate is a sample taken from a random distribution, so some 
knowledge of this distribution and its relationship with the underlying `true' 
value must be available for it to be useful.
Past work in Quantum Monte Carlo has taken this distribution to be Normal, 
usually justified by expressing the estimates as sums of random variables and 
assuming the validity of the Central Limit Theorem (CLT).
It has recently\cite{trail07a} been shown that for the usual implementation of 
QMC (referred to as `standard sampling') this is only partly true for estimates 
of the total energy, and completely untrue for estimates of the (residual) 
variance of the local energy.
These two quantities are the most prominent estimated quantities in Variational 
Monte Carlo (VMC).
For the first of these the deviation of random errors from Normal \textit{may} 
be significant for a finite number of samples in the VMC calculation, with 
outliers occurring.
For the second of these the random error are \textit{not Normal} even in the 
large sample size limit, and large outlier errors are orders of magnitude more 
likely than the CLT suggests.
Such non-Normal distributions of errors are a direct consequence of the 
presence of singularities in the sampled quantities at the nodal surface.
These singularities may not easily be prevented, and have been found to result 
in the failure of the CLT for estimates of many physical expectation values 
sought using QMC methods.

In what follows a new sampling strategy, referred to as `residual sampling', is 
developed that reintroduces the CLT in its strongest form for estimates of the 
total energy and (residual) variance of the local energy.
The paper consists of 6 sections.
In section \ref{sec:res} the new sampling strategy is described.
Sections \ref{sec:toten} and \ref{sec:resvar} describe the construction of 
estimates of the total energy and residual variance within this sampling 
strategy, and derive the distribution of random errors and confidence intervals 
for the estimates.
Section \ref{sec:gensamp} considers the general conditions that a sampling 
strategy must satisfy in order for the CLT to hold for a given estimated 
quantity, so justifying the choice of sampling strategy.
Analytical results, or numerical results for the example case of an isolated 
all-electron carbon atom are presented in each section as appropriate.
Section \ref{sec:otherestimates} considers how an estimate/sampling strategy 
combination may be chosen such that the CLT is valid for an estimate of a 
physical quantity of interest, and the example of the electronic kinetic energy 
is considered.
Section \ref{sec:conc} concludes the paper.

Before commencing we note that this paper is the second of two closely related 
papers.
The preceding paper, \cite{trail07a}, develops the statistical description of 
the random error inherent in QMC, and derives the deficiencies of the standard 
sampling method.
In the current paper, new sampling strategies are developed, together with an 
analysis of the accompanying random errors in estimates.
This provides a method for avoiding the deficiencies of standard sampling by 
controlling the random error and introducing a valid CLT for an estimate of 
interest.

\section{General sampling in VMC, and a new sampling strategy}
\label{sec:res}

Generally, VMC involves generating a statistical estimates of the expectation 
values of an operator of the form 
\begin{equation}
G = \frac{ \langle \psi | \hat{g} | \psi \rangle }
         { \langle \psi | \psi \rangle }.
\end{equation}
Expressing this in terms of the statistical expectation of a function 
$G_L=\psi^{-1}\hat{g} \psi$ sampled over a random distribution of $3N$ 
dimensional vectors, $\text{\sffamily\bfseries R}$, with Probability Density 
Function (PDF) $P(\mathbf{R})$, gives
\begin{equation}
G = \frac{ \mathbb{E}\left[ G_L\psi^2/P;P \right] }
         { \mathbb{E}\left[    \psi^2/P;P \right] },
\end{equation}
where $\mathbb{E}\left[ x ; P \right] = \int x P d\mathbf{R}$ is the definition 
of the expectation.
The function $G_L(\mathbf{R})$ is the `local value' of the operator/trial 
wavefunction combination.
This is true for any general distribution, $P$.

This can also be expressed as statistical estimates constructed from samples 
taken from $P$.
Introducing the notation $\mathsf{A}_r \left[ f \right]$ for an estimate of 
$f$ constructed using $r$ samples, gives
\begin{equation}
\mathsf{A}_r \left[ G \right] = 
    \frac{ \mathbb{E}\left[ G_L\psi^2/P;P \right] + \mathsf{Y}_r }
         { \mathbb{E}\left[    \psi^2/P;P \right] + \mathsf{X}_r }
                                = G + \mathsf{W}_r ,
\end{equation}
where $\mathsf{W}_r$, $\mathsf{Y}_r$, and $\mathsf{X}_r$ are random error 
variables.
The random variable $\mathsf{W}_r$ is not normal, but $\mathsf{Y}_r$ and 
$\mathsf{X}_r$ may be, and may be correlated to some degree.

The `standard sampling' solution is to choose $P=\lambda\psi^2$, with $\lambda$ 
an unknown normalisation constant, so that 
\begin{equation}
\mathsf{A}_r \left[ G \right] = 
    \frac{1}{r} \sum G_L(\text{\sffamily\bfseries R}_n)
                                = G + \mathsf{Y}_r ,
\end{equation}
and $\mathsf{X}_r=0$.
As has previously been shown\cite{trail07a}, singularities in $G_L$ can easily 
prevent the distribution of $\mathsf{Y}_r$ from being Normal by invalidating 
the CLT.
Although standard sampling provides the simplest analytic form for a MC 
estimate, there is nothing to suggest that it is optimum for controlling the 
statistical error in $\mathsf{A}_r \left[ G \right]$.

Returning to general sampling complicates the analysis, but provide a means of 
influencing the random error present in estimated quantities since the 
distribution of the random error, $\mathsf{W}_r$, is influenced by the choice 
of sampling distribution, $P$.

Writing the general sampling distribution as
\begin{equation}
P = \lambda \frac{\psi^2}{w},
\end{equation}
where $\lambda$ is an unknown normalisation factor, provides the estimate of 
$G$ in the more concise form
\begin{equation}
\mathsf{A}_r \left[ G \right] = 
    \frac{ \mathbb{E}\left[ wG_L;P \right] + \mathsf{Y}_r }
         { \mathbb{E}\left[    w;P \right] + \mathsf{X}_r }.
\end{equation}

In order to control the statistics of estimates of the total energy and 
(residual) variance, we begin by introducing the local energy, 
$E_L=\psi^{-1}\hat{H}\psi$, defined in terms of the Hamiltonian operator, 
$\hat{H}$.
We then limit ourselves to weights that are functions of the local energy, 
$w(E_L)$, and to operators of the form $\hat{g}=f(\hat{H})$.
Expectation values of this operator, $F$, then have MC estimates given by
\begin{equation}
\mathsf{A}_r\left[ F \right] =
     \frac{\sum_{n=1}^r w(\mathsf{E}_n) f_L( \mathsf{E}_n ) }
          {\sum_{n=1}^r w(\mathsf{E}_n)                     },
\;\;\;\;\;\;\;\; P(E)=P_{\epsilon}(E)
\end{equation}
where $\mathsf{E}_n$ is the $n^{th}$ independent identically distributed (IID) 
random variable defined as the sample local energy at $\mathsf{R}_n$, and 
distributed as
\begin{eqnarray}
P_{\epsilon}(E)&=& 
    \frac{\lambda}{w(E)} \int_{\partial} 
    \frac{\psi^2}{\left| \nabla_{\mathbf{R}} E_L \right|} 
    d^{3N-1}\mathbf{R} \nonumber \\
               &=& \frac{\lambda'}{w(E)} P_{\psi^2}(E) ,
\label{eq:1}
\end{eqnarray}
where $\lambda'$ is a further unknown normalisation constant, and the integral 
is taken over a $3N-1$ dimensional surface of constant local 
energy\cite{trail07a}.
In the last line, $P_{\psi^2}(E)$ is the distribution of local energies that 
occurs for standard sampling.

Note that $w(E)=1$ results in standard sampling, with the $E^{-4}$ 
leptokurtotic tails for $P_{\epsilon}(E)$, and the resulting CLT issues for VMC.
The essential feature of this approach is that different choices of weight 
function, $w(E)$, provide different estimators for $F$, with a different 
distributions of random error in the estimates.

`Residual sampling' is defined by choosing the weight function to take the 
particular form
\begin{equation}
w(E)= \frac{\epsilon^2}{(E-E_0)^2 + \epsilon^2 },
\label{eq:2}
\end{equation}
where $(E_0,\epsilon)$ are parameters that influence the random error in the 
estimate.
Equation~(\ref{eq:2}) may be interpreted as interpolating between a perfect 
sampling of the numerator and denominator of an estimate of the residual 
variance.
This weight function ensures that, provided $f(E)$ increases quadratically or 
slower in the limit of $E$ approaching infinity from above or below, the 
sampled quantities will be bounded from above and below even in the presence of 
singularities in the local energy.
It is the introduction of this boundary to the sample values that results in 
the re-introduction of the CLT, as described in the next section.
A further significant difference between standard and residual sampling is that 
the former does not sample in the region of the nodal surface, whereas the 
latter does.

From this point on, $w(E)$ refers to Eq.~(\ref{eq:2}), and the accompanying 
distribution of samples in multi-dimensional space is given by
\begin{equation}
P_{\epsilon}(\text{\sffamily\bfseries R}) = 
    \lambda \psi^2(\text{\sffamily\bfseries R} ) / 
    w( E_L(\text{\sffamily\bfseries R}) ) .
\end{equation}
Sampling and estimation using this distribution is straightforward to implement 
in standard Monte Carlo algorithms by using the new distribution at each 
Metropolis step, and by including $w(E)$ when evaluating estimates of 
expectation values.

Values are required for $(E_0,\epsilon)$ to define the sampling strategy, but 
only influence the distribution of random errors in the estimate.
Optimum values (in the sense of resulting in the smallest random error) exist 
and may be sought for a given calculation, but roughly speaking a good choice 
of $E_0$ can be expected to be an approximate total energy, and a good choice 
of $\epsilon$ an estimate of the accuracy of $E_0$.

Two limits exist.
For $\epsilon\rightarrow \infty$ the sampling is as for the standard sampling.
For $(E_0,\epsilon)\rightarrow (E_{tot},0)$ (with $E_{tot}$ the expectation 
value of the trial wavefunction/Hamiltonian combination) the sampling is 
perfect for the numerator of the residual variance estimator, and all the 
statistical error is in the denominator.
For any error in $E_0$ and any finite value of $\epsilon$ this sampling scheme 
is somewhere between these two extremes, hence the numerator is sampled more 
efficiently at the cost of introducing more error in the denominator.
Of course this sampling strategy is only of interest if the estimate converges 
to the true value for increasing sample size ($r$), has a controlled error, 
and is insensitive to the values of the parameters $(E_0,\epsilon)$.

Now that the residual sampling strategy is defined, estimates for the total 
energy and residual variance are considered.
These are of interest in their own right, and from the point of view of 
wavefunction optimisation methods.
The next two sections define these estimates, analyse their statistical 
properties, and obtain distributions of the random error in the large $r$ limit.
In addition numerical VMC calculations for an all-electron carbon atom are 
performed using both standard and residual sampling strategies, in order to 
demonstrate the new sampling strategy for a real system.

It should be borne in mind that many statements about standard sampling are not 
true for a more general sampling method.
An important example is that the residual variance that is to be estimated is 
not the second moment of the sampled quantity, and is unrelated to the error in 
the total energy estimate.

\section{Total energy estimates and confidence limits}
\label{sec:toten}

The residual sampling estimate of the total energy takes the form
\begin{equation}
\mathsf{A}_r\left[ E_{tot} \right] =
     \frac{\sum_{n=1}^r w(\mathsf{E}_n) \mathsf{E}_n }
          {\sum_{n=1}^r w(\mathsf{E}_n)              },
\;\;\;\;\;\;\;\; P(E)=P_{\epsilon}(E).
\end{equation}
In the standard sampling limit $P(E)$ possesses $E^{-4}$ 
asymptotes\cite{trail07a}, but for finite $\epsilon$ the $w(E)^{-1}$ term in 
Eq.~(\ref{eq:1}) results in $E^{-2}$ asymptotic tails.

In order to characterise the random error of this estimate, due consideration 
must be taken of the estimate being made up of a quotient of two random 
variables.
Although $w(\mathsf{E}_n)$ and $\mathsf{E}_n$ are causally related there is no 
reason to expect this causal relationship to hold between sums of these random 
variables, hence the numerator and denominator are only partially correlated.
This observation provides the required route to describing the statistics.

Defining 
\begin{equation}
\left( \mathsf{Y}_n, \mathsf{X}_n \right) = 
\left( w(\mathsf{E}_n) \mathsf{E}_n, w(\mathsf{E}_n) \right)
\end{equation}
provides a bivariate random variable with a PDF that is non-zero only on a 
parametric curve.
A normalised sum of these IID random bivariates gives a new bivariate
\begin{equation}
\left( \mathsf{M}_2, \mathsf{M}_1 \right) = 
\left( 
   \frac{1}{r}\sum_{n=1}^{r} \mathsf{Y}_n,
   \frac{1}{r}\sum_{n=1}^{r} \mathsf{X}_n 
\right) ,
\end{equation}
with a PDF, $P_r( \mu_2, \mu_1)$, that can be derived using a standard 
convolution/Fourier transform approach\cite{stroock93}, and limit theorems 
obtained.
Note that $P_r( \mu_2, \mu_1)$ is not limited to a parametric curve in the two 
dimensional space unless $r=1$.
\footnote{For $r$ underlying variables $\{E_n\}$ the distribution of the total 
set of random variables exists in a $2r$ dimensional space, and the PDF of the 
total set of random variables is non-zero only on a $r$ dimensional surface in 
this space.
The distribution of the two sums is then found by integrating over a 
hyper-plane that defines the numerator, and over a hyper-plane that defines the 
denominator, resulting in the bivariate PDF $P_r( \mu_2, \mu_1)$.}

The total energy estimate is then provided by
\begin{equation}
\mathsf{A}_r\left[ E_{tot} \right] = \frac{\mathsf{M}_2}{\mathsf{M}_1},
\end{equation}
and associated confidence limits must be obtained from the bivariate 
distribution of the numerator and denominator in this expression.

\subsection{Distribution of total energy estimates}
The distribution of errors in the estimate is most easily arrived at by 
initially assuming that the bivariate CLT is valid, and then proving that it is 
so.
For a valid bivariate CLT the random bivariate $(\mathsf{M}_2,\mathsf{M}_1)$ 
possesses the PDF \cite{stroock93}
\begin{equation}
P_r\left( y,x \right) = 
\frac{1}{2\pi} \frac{r^{1/2}}{| C |^{1/2}} e^{-q^2/2}
\label{eq:3}
\end{equation}
in the large $r$ limit.
The function $q$ is defined in matrix notation by
\begin{equation}
q^2= r 
\begin{pmatrix} \left( x - \mu_1 \right) \\ 
                \left(y - \mu_2 \right) \end{pmatrix}^{T}
C^{-1}
\begin{pmatrix} \left( x - \mu_1 \right) \\ 
                \left(y - \mu_2 \right) \end{pmatrix} ,
\label{eq:4}
\end{equation}
where $(\mu_2,\mu_1)=(\mathbb{E}[wE],\mathbb{E}[w])$, and $C$ is the covariance 
matrix defined by the elements
\begin{equation}
c_{ij} = \mathbb{E}\left[ w^2 E^{i+j-2} \right]
       - \mathbb{E}\left[ w   E^{i-1}   \right]
         \mathbb{E}\left[ w   E^{j-1}   \right],
\label{eq:5}
\end{equation}
with $i$ and $j \in \{1,2\}$.
This is the bivariate CLT.

To demonstrate that the CLT is valid it is sufficient to show that all of the 
co-moments of the original distribution exist
\footnote{The existence of all moments is sufficient to show that a 
Gram-Charlier expansion of the characteristic function exists, but not that an 
Edgeworth series expansion exists. It follows from this that the series 
expansion of the characteristic function has a finite radius of convergence 
about the origin, and consequently that there are no asymptotic power laws 
present in the corresponding PDF.}
, or that
\begin{equation}
{\mathcal V}^{m,n}= \mathbb{E}\left[ \left(wE\right)^m \left(w\right)^n \right]
\end{equation}
exists for all non negative $m$ and $n$.
Since the integrand is finite for all $E$, and the asymptotes of $w(E)$ and the 
sampling distribution are known, it follows that the inequality
\begin{eqnarray}
{\mathcal V}^{m,n} 
&<& \int_{-\infty}^{\infty} |P_{\epsilon} w^{m+n} E^m| dE  \nonumber \\
&<& \alpha \int_{-\infty}^{\infty}  \frac{1}{1+|E|^{2n+m+2}} dE
\end{eqnarray}
is true for some finite $\alpha$.
Performing the integral explicitly gives 
\begin{equation}
{\mathcal V}^{m,n} < \frac{2\pi\alpha}{2n+m+2} 
\csc\left( \frac{\pi}{2n+m+2} \right),
\end{equation}
and hence ${\mathcal V}^{m,n}$ is finite for all non negative $m$ and $n$.

This demonstrates that the bivariate CLT is valid, and in addition that 
asymptotic power law behaviour does not occur in the PDF of the random variable 
$(\mathsf{M}_2,\mathsf{M}_1)$ for finite $r$.\cite{gnedenko68}

Now that the validity of the bivariate CLT is established, the distribution of 
the quotient of the two random variables must be considered in order to 
characterise the error in the total energy estimate.
Two approaches to this problem suggest themselves.
The most direct route is to extract the PDF of the quotient directly from the 
bivariate Normal distribution.
An alternative approach is to define a $2$-dimensional confidence region for 
the bivariate distribution.
Both are examined here, with the second proving to be the most appropriate.

A PDF of the quotient is defined in terms of the bivariate PDF via the standard 
formula\cite{curtiss41}
\begin{equation}
P_r(u)= -\int_{-\infty}^{0} x P_r( y=ux, x)  dx 
        +\int_{0}^{+\infty} x P_r( y=ux, x)  dx.
\end{equation}
Evaluating this explicitly using Equations~(\ref{eq:3},\ref{eq:4},\ref{eq:5}), 
and taking the large $r$ limit a second time gives
\begin{eqnarray}
 P_r\left( u \right)&=&
\frac{r^{1/2}}{\sqrt{2\pi}} 
\left|
\frac{
\left( c_{11}\mu_2-c_{12}\mu_1 \right)u + \left( c_{22}\mu_1-c_{12}\mu_2 \right)
}{
\left( c_{11}u^2-2c_{12}u+c_{22} \right)^{3/2}
}
\right|
\nonumber \\ & & \times
\exp{ \left[
-\frac{r}{2}
\frac{ \left( \mu_2-\mu_1u \right )^2           }
     { \left( c_{11}u^2-2c_{12}u+c_{22} \right) }
\right] },
\end{eqnarray}
hence the distribution of quotients is clearly not Normal in the large $r$ 
limit, even though $P_r(\mu_2,\mu_1)$ does approach a bivariate Normal 
distribution.
However, the width of this distribution scales as $r^{-1/2}$ in the same manner 
as a Normal distribution, and for 
$(c_{11},c_{12},\mu_1) \rightarrow (0,0,1)$ this distribution of total
energy estimates approaches a Normal distribution with higher power co-moments
becoming undefined in the limit.

For the general covariance matrix the asymptotic behaviour in $u$ is given by 
\begin{equation}
\lim_{|u|\rightarrow \infty} P_r\left( u \right)=
\frac{r^{1/2}}{\sqrt{2\pi}} 
\left|
\frac{c_{11} \mu_2-c_{12} \mu_1}{c_{11}^{3/2}}
\right|
\exp{ \left[ -\frac{r}{2} \frac{ \mu_1^2 }{c_{11}} \right] }
\frac{1}{u^2},
\label{eq:6}
\end{equation}
hence the distribution of total energy estimates possesses neither a mean or a 
variance.
At first this seems like a serious problem, but it turns out to be irrelevant 
for two reasons.

The magnitude of the power law tails in Eq.~(\ref{eq:6}) decreases 
exponentially as the number of sample points increases, which means that for 
any reasonable set of parameters (and for a wide range of unreasonable 
parameters) the chance of a sample point appearing in these $u^{-2}$ tails is 
vanishingly small.
A typical numerical value for the coefficient of $u^{-2}$ in the asymptotic 
form for calculations actually carried out is $\sim 10^{-4182}$.
In addition the weight, $w(E)$, falls within the closed interval 
$ 0 < \mathsf{X}_n \leq 1$, and $\mathsf{Y}_n$ is also bounded, hence for 
finite sampling these tails do not actually occur.
In effect the deviation of the finite $r$ distribution from the large $r$ limit 
conspires to remove these undesirable tails.

\begin{figure}[t]
\includegraphics{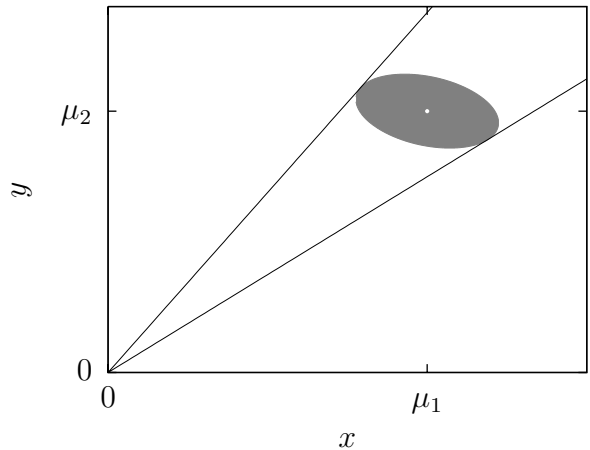}
\caption{\label{fig1}
Figure shows confidence regions defined for a bivariate Normal distribution, 
$P_r(y,x)$, in order to obtain confidence intervals for ratios of the two 
associated random variables.
The grey ellipse follows a line of constant $P_r$, and the straight lines 
enclose a `wedge' that contains lines of gradient $y/x$ with probability 
$\alpha$ (see main text).
}
\end{figure}

The analytic form given above is not the most elegant approach to visualising 
the distribution of gradients.
Confidence intervals for the estimate are more clearly defined directly from 
the bivariate normal distribution by generalising the one dimensional 
confidence interval to a two dimensional confidence region in the space of the 
bivariate PDF.
To achieve this the approach of Fieller\cite{luxburg04} is adopted, and is best 
described geometrically (see Fig.~\ref{fig1}).

An ellipse of constant probability density is defined via a new parameter 
$q_0$ and the equation
\begin{equation}
q_0^2(\alpha_{ellipse})=r
\begin{pmatrix} \left( x - \mu_1 \right) \\ 
                \left(y - \mu_2 \right) \end{pmatrix}^{T}
C^{-1}
\begin{pmatrix} \left( x - \mu_1 \right) \\ 
                \left(y - \mu_2 \right) \end{pmatrix} ,
\end{equation}
which defines an elliptical probability region that contains 
$(\mathsf{M}_2,\mathsf{M}_1)$ with probability $\alpha_{ellipse}$.

A `wedge' is then defined as the region between two straight lines that pass 
through the origin and are tangential to this ellipse of constant probability 
density.
The region contained inside this wedge then defines a second confidence region, 
that contains $(\mathsf{M}_2,\mathsf{M}_1)$ with probability $\alpha$.
Fieller's theorem essentially provides $q_0$ as a function of one variable, 
either $\alpha_{ellipse}$ (via the Hotelling's $T^2$-distribution in the large 
$r$ limit) or $\alpha$ (via the Student's t-distribution in the large $r$ 
limit).
The second of these, $q(\alpha)$, provides a confidence interval for the total 
energy estimate from the confidence wedge, since a fraction $\alpha$ of 
$(\mathsf{M}_2,\mathsf{M}_1)$ provide total energy estimates that fall between 
the bounding lines of the wedge.

Solving for the gradient at the boundaries of the $\alpha$ confidence wedge 
gives
\begin{equation}
l_l < \mathsf{A}_r\left[E_{tot}\right] < l_u \; 
\text{with confidence} \; \alpha ,
\label{eq:7}
\end{equation}
where $l_{u,l}$ are the gradients of wedge boundaries and are given by
\begin{widetext}
\begin{equation}
l_{l,u} = \frac{
\left( r\mu_1.\mu_2-q_0^2c_{12} \right) \pm 
   \sqrt{\left( r\mu_1.\mu_2-q_0^2c_{12} \right)^2 - 
         \left( r\mu_1^2    -q_0^2c_{11} \right)
         \left( r\mu_2^2    -q_0^2c_{22} \right)
        } }{ r\mu_1^2-q_0^2c_{11} },
\label{eq:8}
\end{equation}
and 
\begin{equation}
q_0(\alpha)=\sqrt{2} \textrm{ erf}^{-1}\left( \alpha \right).
\label{eq:9}
\end{equation}
\end{widetext}

For this confidence interval to be finite the ellipse must not cross the $x=0$ 
line since, if it does, the confidence interval may be 
$\mathsf{A}_r\left[E_L\right] > l_u, \mathsf{A}_r\left[E_L\right] < l_l$, 
or even $-\infty < \mathsf{A}_r\left[E_L\right] < \infty$ 
(these two cases are referred to as `exclusive unbounded' and 
`completely unbounded' respectively, with the usual case 
`bounded'\cite{luxburg04}).
A check for whether these `unbounded boundaries' occur is straightforward to 
implement, and is far from being satisfied for systems of interest.
In addition, finite sample size and bounded samples ensure that the unbounded 
cases never occur for the actual (finite $r$) distribution of errors.

The magnitude of the confidence interval scales as $r^{-1/2}$.
It is not immediately apparent what type of estimate is provided by this 
quotient of sample means.
It is a statistical estimate, as in the limit of increasing $r$ it approaches 
the true total energy, however, it is not an unbiased estimate, as its 
distribution has no mean.
In fact no unbiased estimate of the quotient exists, since the mean of a 
quotient of random variables is not equal to the quotient of the mean of the 
random variables.
Equation~(\ref{eq:7}) provides a `central' estimate, in that the probability 
that a sampled estimate value is higher than the true total energy is equal to 
the probability that a sample estimate value is less than the true total 
energy\cite{luxburg04}.

\subsection{Analysis of data}
In this section calculated total energies and confidence limits for an isolated 
all-electron carbon atom are considered, both using standard sampling and 
residual sampling.

A numerical Multi-Configuration-Hartree-Fock calculation was performed to 
generate a multideterminant wavefunction consisting of $48$ Slater determinants 
(corresponding to 7 configuration state functions (CSF)) using the ATSP2K code 
of Fischer \textit{et al.}\cite{fischer07}.
Further correlation was introduced via a $83$ parameter Jastrow 
factor\cite{drummond04}, and a $130$ parameter backflow 
transformation\cite{rios06}.
This $219$ parameter trial wavefunction was optimised using a standard variance 
minimisation method\cite{casino06}, resulting in $E_{VMC}=-37.8344(2)$ a.u., 
compared with the `exact'\cite{chakravorty93} result of $-37.8450$ a.u.
Of those trial wavefunctions that can practically be constructed and used in 
QMC this may be considered to be accurate, and reproduces $93.2 \%$ of the 
correlation energy at VMC level.
Unless otherwise stated the parameters $(E_0,\epsilon)$ are taken to be the 
estimated total energy and variance of the local energy taken from a small 
standard sampling calculation.
This choice is justified in what follows.

The analysis of the sampled local energies uses the formulae derived above, 
with the expectation integrals replaced by the normalised sums of samples that 
are the standard unbiased estimates.
The sampled estimate of the quantity $x$ is denoted $\widehat{x}$, and sample 
estimates of the bivariate mean and covariance matrix were calculated.
The primary aim of analysing the data is to characterise the statistics of the 
random error in sample estimates for both residual and standard sampling.
Generating $10^6$ local energy samples, breaking this set of samples into 
subsets of various sizes and analysing each of the subsets individually 
provides independent sample estimates for the total energy and variance, and 
these are then analysed as a set of samples from the underlying distribution, 
$P_r$.

Within residual sampling the sample estimate of the bivariate mean obtained 
from $r$ samples is 
\begin{equation}
\left( \widehat{\mu}_2 , \widehat{\mu}_1 \right) = 
\left( 
   \frac{1}{r}\sum_{n=1}^{r} {w_nE_n},\frac{1}{r}\sum_{n=1}^{r} {w_n} 
\right),
\end{equation}
and the sample estimate of the covariance matrix elements take the form
\begin{eqnarray}
\widehat{c}_{22}&=& \frac{1}{r-1} 
    \sum_{n=1}^{r} \left( w_nE_n - \widehat{\mu}_2 \right)^2    \nonumber \\
\widehat{c}_{12}&=& \frac{1}{r-1} 
    \sum_{n=1}^{r} \left( w_nE_n - \widehat{\mu}_2 \right)
                   \left( w_n - \widehat{\mu}_1 \right) \nonumber \\
\widehat{c}_{11}&=& \frac{1}{r-1} 
    \sum_{n=1}^{r} \left( w_n - \widehat{\mu}_1 \right)^2.
\end{eqnarray}

These provide an estimated value of the total energy and accompanying 
confidence limits 
\begin{equation}
\widehat{E}_{tot} = \frac{ \widehat{\mu}_2 }{ \widehat{\mu}_1 },
\end{equation}
and 
\begin{equation}
\widehat{l}_l < E_{tot} < \widehat{l}_u \; \text{with confidence} \; \alpha ,
\end{equation}
with the limits given by
\begin{widetext}
\begin{equation}
\widehat{l}_{u/l} = 
\frac{
\left( r\widehat{\mu}_1.\widehat{\mu}_2-q_0^2\widehat{c}_{12} \right) \pm 
   \sqrt{
         \left( 
                r\widehat{\mu}_1.\widehat{\mu}_2-q_0^2\widehat{c}_{12}
         \right)^2 - 
         \left( r\widehat{\mu}_1^2    -q_0^2\widehat{c}_{11} \right)
         \left( r\widehat{\mu}_2^2    -q_0^2\widehat{c}_{22} \right)
        } }{ r\widehat{\mu}_1^2-q_0^2\widehat{c}_{11} } ,
\end{equation}
and $q_0$ a function of the required confidence interval via Eq.~(\ref{eq:9}).
\end{widetext}
If required, further information on the deviation of this distribution from the 
large $r$ limit is available from statistical estimates of higher co-moments, a 
fundamentally different situation to the standard sampling case.

\begin{figure}[t]
\includegraphics{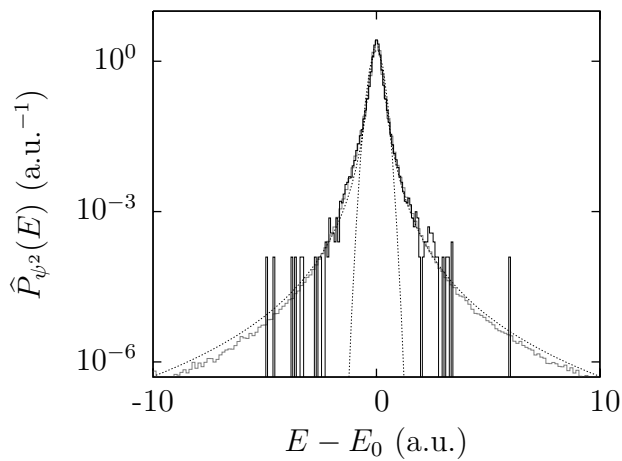}
\caption{\label{fig2}
The seed probability density function estimated by a histogram of $r=10^6$ 
sampled local energies using standard sampling (black) and residual sampling 
(grey).
These are results for an accurate all-electron carbon trial wavefunction, as 
described in the text.
Also shown is the model distribution of Eq.~(\ref{eq:10}) that reproduces the 
mean and variance of the samples.
}
\end{figure}

Figure~\ref{fig2} shows estimates of the seed PDF, $P_{\psi^2}(E)$, constructed 
from taking $10^6$ samples of the local energy, binning these into intervals, 
and normalising\cite{izenman91}.
Estimates are constructed from both standard sampling, for which a weight of 
$1$ per sample is binned into the chosen energy intervals, and residual 
sampling, for which a weight $w(E_n)$ is binned.
In addition the figure shows a `model' distribution of the form
\begin{equation}
p(E)=\frac{\sqrt{2}}{\pi} \frac{\widehat{\sigma}^3}{\widehat{\sigma}^4+
     \left(E-\widehat{E}_{tot}\right)^4},
\label{eq:10}
\end{equation}
with a mean and variance of $\widehat{E}_{tot}$ and $\widehat{\sigma}^2$ whose 
values are obtained from the data using the usual unbiased estimates.
This is chosen as a simple analytic form that reproduces the $E^{-4}$ 
asymptotic behaviour that has been shown to be present in the seed 
distribution\cite{trail07a}.

It is clear that residual sampling takes into account the statistics of the 
local energy for large deviations from the estimated total energy far more 
precisely than standard sampling.
The energy range of the figure is chosen to show the breakdown of standard 
sampling, but for residual sampling the estimated PDF shows the same same 
precision over an interval of around $1000$ a.u.
In addition the expected $E^{-4}$ asymptotic behaviour (and agreement with the 
model distribution) are reproduced by the estimate over this range.
This demonstrates a distinct difference between the two approaches - standard 
sampling does not sample the nodal surface and this results in weak statistical 
convergence to the underlying PDF, whereas residual sampling does sample the 
nodal surface successfully, resulting in a faster statistical convergence to 
the underlying PDF.

\begin{figure}[t]
\includegraphics{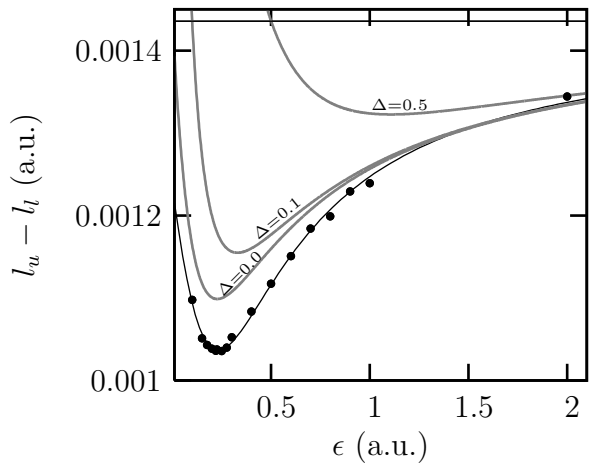}
\caption{\label{fig3}
Confidence limits for estimates of the total energy for residual sampling, as a 
function of $(E_0,\epsilon)$.
Data points (with a fitted Pad\'e form to guide the eye) are calculated for 
$E_0$ taken as the standard sampling estimate of the total energy.
Grey curves are the confidence limits resulting from the model distribution of 
Eq.~(\ref{eq:10}), with $\Delta$ the positive deviation from the exact VMC 
energy.
The horizontal line at $l_u-l_l=0.001436$ a.u. is the standard sampling limit 
corresponding to $\epsilon$ approaching infinity.
}
\end{figure}

Residual sampling requires a choice of parameters to specify the sampling PDF, 
$(E_0,\epsilon)$.
Although the values of these parameters influence only the statistics of the 
random errors in estimates, it is important to examine how variations in these 
parameters change the confidence ranges for estimates.
Figure~\ref{fig3} shows the estimates of $l_u-l_l$ that result from the 
numerical calculations as a function of $\epsilon$.
Each datum was obtained using $r=10^5$ samples, for a range of $\epsilon$ 
values, and for a fixed $E_0=-37.8344$ a.u., the standard sampling total energy 
estimate for the trial wavefunction.
The confidence range possesses a well defined minimum for $\epsilon$ close to 
the standard deviation of $P_{\psi^2}$, and for increasing $\epsilon$ 
approaches the standard sampling limit.
The optimum confidence range (assumed to be at $\epsilon=\widehat{\sigma}$) is 
approximately $75\%$ of that resulting from standard sampling.

Also shown in the figure are the confidence ranges obtained analytically for 
the model distribution of Eq.~(\ref{eq:10}).
The figure shows the same general behaviour for the model and actual 
distribution, with higher accuracy for the actual results.
The confidence range is shown as several functions of $\epsilon$, with $E_0$ 
chosen to overestimate the true mean value (known for the model distribution) 
by an increasing amount, $\Delta$.
The results show that for the model system the presence of an improved 
confidence interval is resilient to the deviations of the parameters $E_0$ and 
$\epsilon$ from their optimum values.

For the model distribution the optimum reduction in the error relative to 
standard sampling is a factor of $0.765$, which occurs for 
$(E_0,\epsilon)=(E_{tot},\sigma)$ in the large $r$ limit.
The results suggest that an inaccurate estimate of $E_{tot}$ can be used for 
$E_0$ (an accuracy of better that $0.5$ a.u. should be sufficient), and that an 
order of magnitude estimate of the variance of the local energy may be used for 
$\epsilon$.
Should this be insufficient it is always possible to optimise the confidence 
interval itself with respect to variations in $(E_0,\epsilon)$.

\begin{figure}[t]
\includegraphics{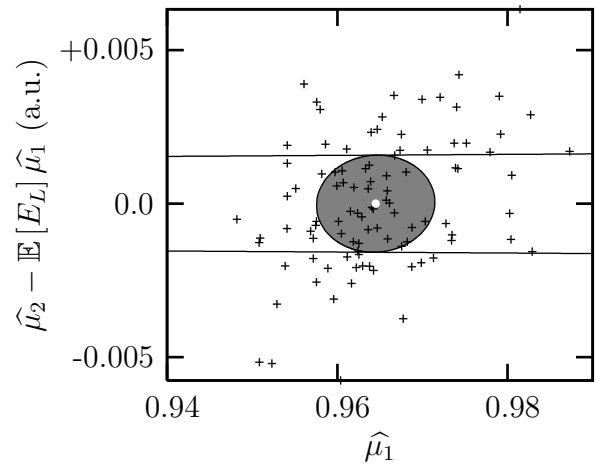}
\caption{\label{fig4}
This figure shows the statistics of estimate values of the total energy.
Scattered points are $100$ estimated values of the means whose quotient 
provides total energy estimates.
The ellipse is the estimated confidence ellipse, and the two straight line 
enclose the estimated confidence wedge described in the main body of the text.
For a valid bivariate CLT, $68.3 \%$ of estimates fall within the confidence 
wedge.
}
\end{figure}

In calculating the confidence intervals in Fig.~\ref{fig3} it is implicitly 
assumed that the large $r$ limit has effectively been reached.
It is desirable to convincingly show that this is in fact the case for the 
example calculation considered here.
First a `big' estimate of the bivariate mean and covariance matrix is 
constructed from the $10^6$ sample local energies.
Then this set of local energy samples is separated into $10^2$ blocks of $10^4$ 
samples, and $10^2$ estimates of the bivariate mean are constructed from these 
blocks of data.

Figure~\ref{fig4} shows the confidence ellipse and confidence wedge of the 
$r=10^4$ estimates predicted using the `big' estimate of the bivariate mean and 
covariance matrix.
In addition the $10^2$ $(\widehat{\mu}_2,\widehat{\mu}_1)$ estimates are also 
scattered over the figure.
Of the sampled bivariates, $62$ fall within the $68.3\%$ confidence wedge, in 
good agreement with the bivariate CLT, and no suspicious outliers occur.
It should be noted that a linear combination of the means is plotted on the 
vertical axis of the figure to make the finite width of the confidence wedge 
visible - otherwise the correlation between the sample means dominates and all 
samples appear to fall on a line passing through the origin and with a gradient 
given by the total energy.

This data supports the suitability of the residual sampling strategy, bivariate 
CLT, and the accompanying interpretation of error.

\begin{figure}[t]
\includegraphics{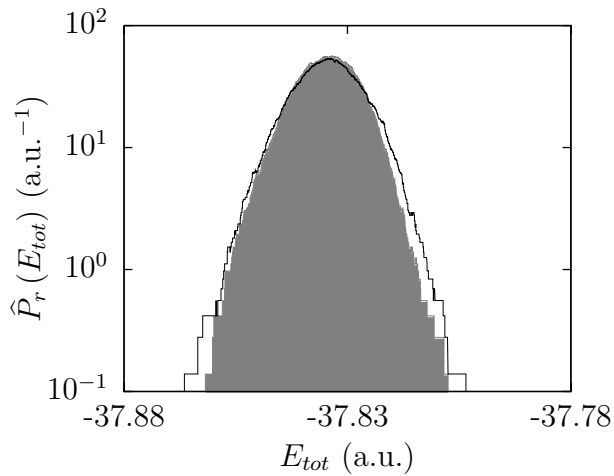}
\caption{\label{fig5}
Estimated PDFs for total energy estimates constructed from different sampling 
strategies.
The unfilled solid curve is for standard sampling, and the grey filled curve 
for residual sampling.
In both cases, a kernel estimate of the PDF was constructed from $10^3$ 
total energy estimates, with each total energy estimate constructed from 
$r=10^3$ samples.
}
\end{figure}

Finally, an estimate of the PDF for total energy estimates is constructed from 
the numerical data, for both standard and residual sampling.
Dividing the $10^6$ samples into $10^3$ blocks of $10^3$ samples provides 
$10^3$ sample estimates of the total energy in each case.
A kernel estimate\cite{izenman91} of the distribution of total energy estimates 
is then constructed using 
\begin{equation}
P_r\left(E\right)=\frac{1}{mh} \sum 
    \Theta \left( \frac{E-\mathsf{A}_{r} \left[ E_{tot} \right]}{h} \right),
\label{eq:11}
\end{equation}
where the kernel, $\Theta$, was chosen to be a centred top-hat function of 
width $1$, $m=10^3$ is the number of estimates, and $h$ is the width parameter, 
chosen heuristically to provide the clearest plot.

The estimated $P_r(E)$ for standard and residual sampling is shown in 
Fig.~\ref{fig5}.
Although $E^{-4}$ asymptotic tails are known to be present in the distribution 
for standard sampling total energy estimates, for this particular calculation 
they are not significant at an achievable statistical resolution.
There is no guarantee that this will be the case for other 
calculations\cite{trail07a}.
For residual sampling the bivariate CLT is valid in its strongest form, hence 
such persistent leptokurtotic tails are guaranteed to be absent.

Assuming the large $r$ limit has been reached, it is apparent that residual 
sampling provides an improved confidence interval ($\sim75\%$ of the standard 
sampling interval), with an estimated total energy of $-37.8344(23)$ a.u. for 
standard sampling, and $-37.8346(16)$ for residual sampling.
To put this another way, residual sampling requires approximately half as many 
samples as standard sampling to achieve a given accuracy.

\section{Residual variance estimates and confidence limits}
\label{sec:resvar}

The residual variance, $V_{\delta^2}$, is defined as the integral of the square 
of the residual associated with the Schr\"odinger equation,
\begin{equation}
V_{\delta^2} =
\frac{ \langle \psi | (\hat{H} - E_G)\cdot(\hat{H} - E_G) | \psi \rangle}
     { \langle \psi | \psi \rangle},
\end{equation}
where $E_G$ may be considered as a variational parameter\cite{nakatsuji04}.
In terms of expectation of functions over the seed distribution of residual 
sampling, $P_{\epsilon}$, this takes the form
\begin{equation}
V_{\delta^2} = 
    \frac{ \mathbb{E}\left[ w\left(E-E_G\right)^2\right] }
         { \mathbb{E}\left[ w \right]                    }.
\label{eq:12}
\end{equation}
The parameter $E_G$ may be varied to minimise the residual variance, or taken 
to be the total energy (the two are equivalent if the expectations in the above 
equation are not estimated).

For standard sampling the CLT is not valid for estimates of the residual 
variance\cite{trail07a}.
This, together with the importance of the residual variance in wavefunction 
optimisation methods, makes the development of an improved residual variance 
estimator desirable.

\subsection{Distribution of residual variance estimates}
Taking $E_{G}$ to be the total energy gives Eq.~(\ref{eq:12}) in the form
\begin{equation}
V_{\delta^2} =
\frac{\mathbb{E} \left[ wE^2 \right]}{\mathbb{E} \left[ w    \right]}
-\left(
\frac{\mathbb{E} \left[ wE   \right]}{\mathbb{E} \left[ w    \right]}
\right)^2,
\end{equation}
with a statistical estimate of this quantity provided by replacing each 
expectation by a normalised sum of samples.

A rigorous treatment of the statistics of this estimate requires a 
generalisation of the bivariate analysis to the trivariate case using
\begin{equation}
\left( \mathsf{A}_n,\mathsf{B}_n,\mathsf{C}_n \right) = 
\left(
w( \mathsf{E}_n ) \mathsf{E}_n^2 ,
w( \mathsf{E}_n ) \mathsf{E}_n   ,
w( \mathsf{E}_n )
\right) ,
\end{equation}
and the accompanying unbiased estimates of the means that form the partially 
correlated random trivariate,
\begin{equation}
\left( \mathsf{M}_2,\mathsf{M}_1,\mathsf{M}_0 \right)=
\left(
\frac{1}{r}\sum_{n=1}^{r} \mathsf{A}_n,
\frac{1}{r}\sum_{n=1}^{r} \mathsf{B}_n,
\frac{1}{r}\sum_{n=1}^{r} \mathsf{C}_n
\right),
\end{equation}
to provide the estimated residual variance as
\begin{equation}
\mathsf{A}_r\left[ V_{\delta^2} \right] =
      \frac{\mathsf{M}_2}{\mathsf{M}_0}-
\left(\frac{\mathsf{M}_1}{\mathsf{M}_0}\right)^2.
\end{equation}
Confidence intervals for this quantity may, in principle, be obtained by an 
analogous route to the bivariate case, by obtaining an (unbiased) estimate of a 
$3 \times 3$ covariance matrix and defining a confidence region in the 
$3$-dimensional space to provide a trivariate CLT and an analogue of Fieller's 
theorem.
This added complexity is not considered to be necessary here.

Instead, $E_G$ is interpreted as a variational parameter which results in an 
estimate of the residual variance that takes a bivariate form, and that 
reproduces standard sampling for $w=1$ and finite $r$.
A random bivariate is defined as
\begin{equation}
\left( \mathsf{Y}_n,\mathsf{X}_n \right) = 
\left(
w( \mathsf{E}_n ) \left( \mathsf{E}_n - E_G \right)^2,
w( \mathsf{E}_n )
\right)
\end{equation}
The associated bivariate
\begin{equation}
\left( \mathsf{M}_2,\mathsf{M}_1 \right)=
\left(
    \frac{1}{r-1}\sum_{n=1}^{r} \mathsf{Y}_n,
    \frac{1}{r}\sum_{n=1}^{r}   \mathsf{X}_n
\right)
\end{equation}
provides the random variables whose quotient is an estimate of the residual 
variance
\begin{equation}
\mathsf{A}_r\left[ V_{\delta^2} \right] =
            \frac{\mathsf{M}_2}{\mathsf{M}_1}.
\end{equation}
The prefactor in the definition of $\mathsf{M}_2$ ensures that the above 
estimate is unbiased for the case of standard sampling.
As for total energy estimates, the bivariate CLT is assumed to be valid in 
order to define the distribution of $(\mathsf{M}_2,\mathsf{M}_1)$, and then 
shown to be valid.

Provided the CLT is valid, the large $r$ PDF takes the form 
\begin{equation}
P_r\left( y,x \right) =
\frac{1}{2\pi}
\frac{r^{1/2}}{ | C |^{1/2} } e^{-q^2/2} ,
\end{equation}
and
\begin{equation}
q^2=r
\begin{pmatrix} \left( x - \mu_1 \right) \\ 
                \left( y - \mu_2 \right) \end{pmatrix}^{T}
C^{-1}
\begin{pmatrix} \left( x - \mu_1 \right) \\ 
                \left( y - \mu_2 \right) \end{pmatrix} .
\end{equation}
The bivariate mean $(\mu_2,\mu_1)$ and covariance matrix, $C$, are defined in 
terms of the supplementary variables 
$\left(x_2,x_1\right)=\left( w\left(E-E_G\right)^2 , w\right)$ by
\begin{equation}
(\mu_2,\mu_1)=(\mathbb{E}[x_2],\mathbb{E}[x_1]),
\end{equation}
and
\begin{equation}
c_{ij} = \mathbb{E}\left[ x_{i}x_{j} \right]
        -\mathbb{E}\left[ x_{i}      \right]
         \mathbb{E}\left[ x_{j}      \right]
\end{equation}
for $i$ and $j \in \{1,2\}$.
This is the bivariate CLT.

To show that this CLT is valid it is sufficient to show that all of the 
co-moments of the original distribution exist.
A general co-moment can be expressed in terms of the weights and energies as
\begin{eqnarray}
{\mathcal V}^{m,n}&=& \mathbb{E}\left[ x_2^m x_1^n \right] \\
&=& \sum_{k=0}^{2m} 
\begin{pmatrix} 2m \\ k \end{pmatrix} 
E_{G}^{2m-k} \mathbb{E}\left[ w^{m+n} E^k \right] ,
\end{eqnarray}
hence it is required to show that $\mathbb{E}\left[ w^{m+n} E^k \right]$ is 
finite for all $m,n$ and $0 \leq k \leq 2m$ (this includes the co-moments 
associated with $E_{tot}$).
Noting that the integrand is finite for all $E$, and possesses asymptotes 
proportional to $E^{k-2-2(m+n)}$ provides the inequalities
\begin{eqnarray}
\mathbb{E}\left[ w^{m+n} E^{k} \right]
&<& \int_{-\infty}^{\infty} |P_{\epsilon} w^{m+n} E^{k}| dE 
\nonumber \\
&<& \alpha \int_{-\infty}^{\infty}  \frac{1}{1+|E|^{2-k+2(m+n)}} dE ,
\end{eqnarray}
for some finite $\alpha$, or that
\begin{equation}
\mathbb{E}\left[ w^{m+n} E^{k} \right] < 
   \frac{2\pi\alpha}{2-k+2(m+n)} \csc\left( \frac{\pi}{2-k+2(m+n)} \right).
\end{equation}
This inequality is valid for all non-negative $m$,$n$ and $0 \leq k \leq 2m$, 
and hence all co-moments exist.
It then follows that the bivariate CLT is valid and no asymptotic power law 
behaviour occurs in the PDF of $(\mathsf{M}_2,\mathsf{M}_1)$.
Converting this bivariate distribution into a description of the statistics of 
the residual variance estimate proceeds exactly as for the total energy 
estimates in the previous section.
All that differs is the definition of the bivariate mean and the covariance 
matrix.

From this point on, and in all numerical results, we choose $E_G=E_{tot}$, with 
$E_{tot}$ taken as the estimate of the previous section.
Any deviation of $E_G$ from the true expectation value of the total energy of 
$\psi$ does not invalidate the variational principle for which the residual 
variance is of interest, but it should be borne in mind that the relatively 
small random variation in $E_G$ is not taken into account in this error 
analysis.

\subsection{Analysis of data}
Returning to the all-electron carbon atom, a VMC estimate of the residual 
variance is required.
The same local energy samples used for the total energy estimates are used to 
construct the residual variance estimates.

First a `central' estimate of the total energy is constructed,
\begin{equation}
\widehat{E}_{tot} = \frac{ \sum_{n=1}^{r} w_nE_n }{ \sum_{n=1}^{r} w_n },
\end{equation}
and this is used to construct an estimate of the mean bivariate 
\begin{equation}
\left( \widehat{\mu}_2 , \widehat{\mu}_1 \right) =
\left(
\frac{1}{r-1}\sum_{n=1}^{r} w_n\left(E_n - \widehat{E}_{tot}\right)^2 ,  
\frac{1}{r}  \sum_{n=1}^{r} w_n
\right)
\label{eq:13}
\end{equation}
and covariance matrix elements 
\begin{eqnarray}
\widehat{c}_{22}&=& \frac{1}{r-1} \sum_{n=1}^{r} 
 \left[ w_n(E_n - \widehat{E}_{tot})^2 - \widehat{\mu}_2 \right]^2 \nonumber \\
\widehat{c}_{12}&=& \frac{1}{r-1} \sum_{n=1}^{r} 
 \left[ w_n(E_n - \widehat{E}_{tot})^2 - \widehat{\mu}_2 \right].
 \left[ w_n - \widehat{\mu}_1                            \right]   \nonumber \\
\widehat{c}_{11}&=& \frac{1}{r-1} \sum_{n=1}^{r}
 \left[ w_n - \widehat{\mu}_1                            \right]^2.
\label{eq:14}
\end{eqnarray}

Equations~(\ref{eq:13},\ref{eq:14}) provide the sample estimate of the residual 
variance as
\begin{equation}
\widehat{V}_{\delta^2} =
\frac{ \widehat{\mu}_2 }{ \widehat{\mu}_1 } ,
\end{equation}
with 
\begin{equation}
\widehat{l}_l < V_{\delta^2} < \widehat{l}_u \; 
\text{with confidence} \; \alpha ,
\end{equation}
and $\widehat{l}_{u/l}$ defined in terms of the new 
$(\widehat{\mu}_2,\widehat{\mu}_1)$ and $C$ using the Fieller's theorem and 
Eq.~(\ref{eq:8}).
As before, further information on the deviation of this distribution from the 
large $r$ limit is available from estimates of higher moments.

\begin{figure}[t]
\includegraphics{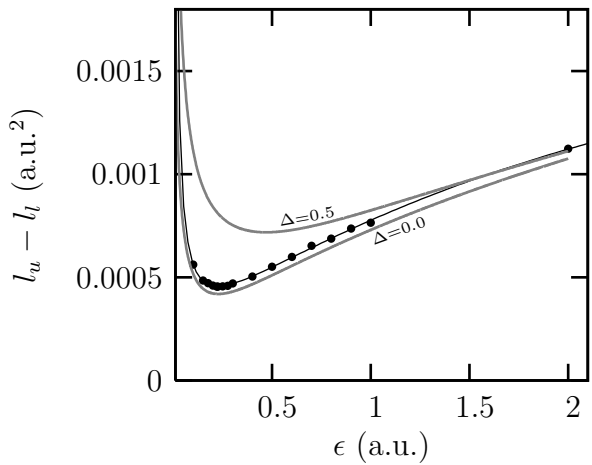}
\caption{\label{fig6}
Confidence limits for estimates of the residual variance for residual sampling, 
as a function of $(E_0,\epsilon)$.
Data points (with a fitted Pad\'e form to guide the eye) are calculated for 
$E_0$ taken as the standard sampling estimate of the total energy.
Grey curves are the confidence limits resulting from the model distribution of 
Eq.~(\ref{eq:10}), with $\Delta$ the positive deviation from the exact total 
energy.
The standard sampling limit for this quantity that corresponds to $\epsilon$ 
approaching infinity is not defined.
}
\end{figure}

Results for the all-electron carbon atom are now considered in the same manner 
as for the total energy estimates of the previous section, and for the same 
reason.
Beginning with the influence of the sampling parameters, $(E_0,\epsilon)$, on 
the statistical error, Fig.~\ref{fig6} shows estimates of $l_u-l_l$ that result 
from the numerical calculation for a range of values of $\epsilon$.
Each datum was obtained using $r=10^5$ samples, and for a fixed $E_0=-37.8344$ 
a.u., the standard sampling total energy estimate for the trial wavefunction.
As for the total energy estimate, the confidence range possesses a well defined 
minimum for $\epsilon$ close to the standard deviation of $P_{\psi^2}$.
However, unlike the total energy estimate, this is not a finite reduction of 
the CLT confidence range of standard sampling, since for standard sampling the 
CLT confidence range is not defined.
In other words $l_u-l_l$ is unbounded as $\epsilon$ increases, and no sample 
estimate of the standard sampling confidence interval is shown as such a 
quantity does not exist.

The figure also shows the confidence ranges resulting from the model seed 
distribution (Eq.~(\ref{eq:10})), obtained analytically and plotted as 
functions of $\epsilon$ for $E_0$ chosen to overestimate the true mean value 
(known for the model distribution) by $\Delta$.
The analytic form shows no upper bound, as expected, and suggests that the 
usefulness of the confidence range is resilient to the deviations of the 
parameters $E_0$ and $\epsilon$ from their optimum values.
Given that no `standard sampling confidence range' exists, the case for 
improved accuracy for residual sampling is stronger than for the total energy 
estimate.
Parameter values may be chosen by the same criteria suggested for total energy 
estimates, or by minimising the confidence interval itself.

\begin{figure}[t]
\includegraphics{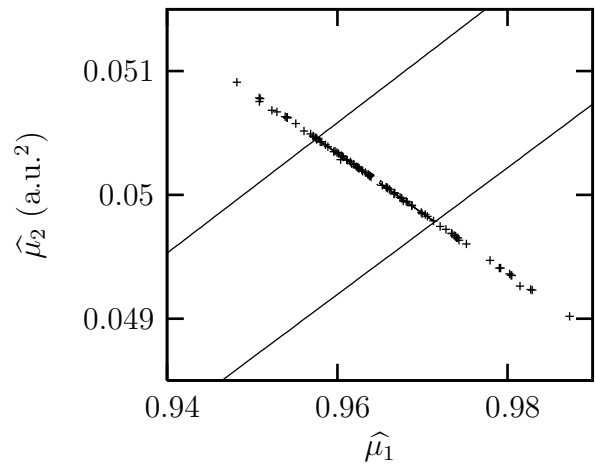}
\caption{\label{fig7} 
This figure shows the statistics of estimate values of the residual variance.
Scattered points are $100$ estimated values of the means whose quotient 
provides residual variance estimates.
The two straight line enclose the estimated confidence wedge described in the 
main body of the text.
For a valid bivariate CLT, $68.3 \%$ of estimates fall within the confidence 
wedge.
}
\end{figure}

To justify the validity of having reached the large $r$ limit with real 
numerical results, and the related validity of the bivariate CLT, the $10^6$ 
sample local energies were used to generate $10^2$ estimates of the bivariate 
mean made up of $r=10^4$ samples each, and an estimate of the distribution that 
these are sampled from.
The quantity $E_G$ was defined as the estimate of the total energy defined in 
section \ref{sec:toten}, evaluated separately for each block.
Figure~\ref{fig7} shows a confidence wedge predicted for the estimates 
constructed from the sample covariance and mean taken from all the samples, and 
also shows the $10^2$ $(\widehat{\mu}_2,\widehat{\mu}_1)$ estimates scattered 
over the figure.
Of the sampled bivariates, $66$ fall within the $68.3\%$ confidence wedge, in 
agreement with the bivariate CLT, and no suspicious outliers occur.
This also justifies the bivariate interpretation of the residual variance 
estimate by showing that the statistical variation in $E_G$ is not significant.
Note that the degree of correlation (although not complete) prevents the 
confidence ellipse being visible.
This data supports the suitability of residual sampling, the bivariate CLT, and 
the accompanying interpretation of error for obtaining estimates of the 
residual variance.
This is fundamentally different to the standard sampling case, where no CLT is 
valid and the statistical error is uncontrolled.

\begin{figure}[t]
\includegraphics{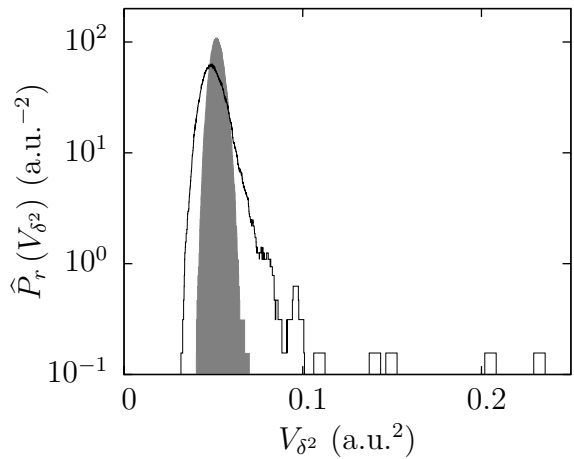}
\caption{\label{fig8}
Estimated PDFs for residual variance estimates constructed from different 
sampling strategies.
The unfilled solid curve is for standard sampling, and the grey filled curve 
for residual sampling.
In both cases, a kernel estimate of the PDF was constructed from $10^3$ 
residual variance estimates, with each residual variance estimate constructed 
from $r=10^3$ samples.
}
\end{figure}

Finally, a kernel estimate to the PDF of the residual variance estimate is 
constructed for both standard and residual sampling in order to compare the 
distributions of error that result in the two cases.
The estimated PDFs where constructed by dividing $10^6$ local energy samples 
into $10^3$ blocks of $10^3$ samples, constructing a residual variance estimate 
for each block (using a block by block total energy estimate), and then 
constructing a kernel estimate using Eq.~(\ref{eq:11}).
Figure~\ref{fig8} show the resulting estimated PDFs.

The estimated standard sampling distribution clearly demonstrate the invalidity 
of the CLT, leptokurtotic tails, and accompanying outliers predicted for 
standard sampling in a previous paper\cite{trail07a}.
The estimated residual sampling distribution reflects the error analysis given 
earlier in this section, providing numerical evidence that the large $r$ limit 
of the bivariate CLT has been reached.

On comparing the properties of the two distributions, two main points suggest 
themselves.
Due to the presence of power law tails for standard sampling, it provides a far 
wider distribution and is more vulnerable to outliers than residual sampling.
In addition, for increasing $r$, the statistical spread of estimates scales as 
$r^{-1/3}$\cite{trail07a} and $r^{-1/2}$ for standard and residual sampling 
respectively, hence standard sampling becomes even less accurate relative to 
residual sampling as the number of samples increases.

Essentially this data tells us that the random error in estimates of the 
residual variance is very different for standard and residual sampling.
The CLT fails for standard sampling, but is reintroduced for residual sampling, 
so residual sampling provides a confidence interval for the residual variance, 
whereas standard sampling does not.
In addition the data suggest that the large $r$ limit is easily reached for 
practical sample sizes.
The model seed distribution of Eq.~(\ref{eq:10}) and the numerical data for the 
carbon atom suggests a standard sampling error one to two orders of magnitude 
larger than for residual sampling for $r=10^3$, and this ratio increases as 
$r^{1/6}$.

\section{General sampling and moments of seed distribution}
\label{sec:gensamp}
The analysis given above has involved only a particular sampling/weighting 
function combination, referred to as residual sampling.
A more general sampling function is now considered in order to show how the 
presence of $E^{-4}$ asymptotic behaviour in the `standard' distribution of 
local energies $P_{\psi^2}$ limits the quantities that may be estimated, and 
the statistics of the random errors in those quantities that can be estimated.

The influence of the chosen weighting/sampling functions on the applicable 
limit theorems can be characterised by its asymptotic behaviour, specifically 
by the inverse power law behaviour of the weight function as singularities in 
the local energy are approached.
A large $E$ power law behaviour of $w \propto |E|^{-p}$ is taken for the 
weight, and used to estimate the $q^{th}$ physical moment of the seed 
distribution,
\begin{equation}
m_{q}=\int_{-\infty}^{\infty} P_{\psi^2} E^{q} dE.
\end{equation}
The limit theorem valid for this moment will also be valid for the expectation 
of any function of $E$ that increases as $E^q$ in the large $|E|$ limit.

The distribution of an estimate of this moment will satisfy the CLT in its 
strongest form if all of the co-moments for the sampling strategy characterised 
by $w(E)$ exists, that is if 
\begin{equation}
\mathcal{V}^{m,n}=\mathbb{E}\left[ (wE^q)^m(w)^n \right],
\end{equation}
exists.
This is the case if the inequalities
\begin{eqnarray}
\mathcal{V}^{m,n} 
  &\leq & \int_{-\infty}^{\infty} | P_{\epsilon} w^{m+n} E^{qm}| dE
\nonumber \\
  &<    & \alpha \int_{-\infty}^{\infty} \frac{1}{
                         1 + |E|^{  4-p-qm+p(m+n)} } dE
\end{eqnarray}
are satisfied for all non-negative $m$,$n$ and some finite $\alpha$.
The integral on the RHS is finite provided that
\begin{equation}
n > 1-\frac{3}{p} + m \left(\frac{q}{p}-1\right),
\end{equation}
which is true for all non-negative $m$,$n$ provided that
\begin{equation}
p<3 \; \text{and} \; q \leq p.
\end{equation}

If this pair of inequalities is satisfied then the least general version of the 
bivariate CLT (that provides the strongest limits on the deviation from a 
Gaussian distribution) is valid for the estimated moment.
This is the most desirable case, and precludes the presence of power law tails 
for finite $r$.
Note that this inequality demonstrates that it is not possible to estimate 
$3^{rd}$ moments or higher of the $P_{\psi^2}$, which is not surprising given 
that these integrals are not defined.
The value of $p$ is an exclusive upper limit on the moments that can be 
estimated with strong limits on their statistical error, and cannot be greater 
than or equal to $3$.

The most general version of the bivariate CLT that provides no limit on the 
deviation from a Gaussian PDF for finite $r$ is the bivariate form of the 
Lindeberg theorem\cite{stroock93}.
For this theorem to hold requires only the $1^{st}$ and $2^{nd}$ order 
co-moments of the estimate to exist, resulting in the weaker limits
\begin{equation}
p<3 \; \text{and} \; q \leq \frac{3+p}{2}.
\end{equation}
So there is a small range of $q$ values between the existence of all moments 
and the complete invalidity of the bivariate CLT where power law tails will 
persist into the distribution of statistical errors.
No integer $q$ falls in this region.

The $q$,$p$ values for which all moments exist tells us that the CLT with the 
strongest limits on finite sample error is valid for estimates of all the 
expectations that exist for the trial wavefunction.
Standard sampling does not provide this ideal strategy of sampling and 
estimation, and many of the expectations that exist have estimates that either 
satisfy the CLT with the weakest limits on the finite sample error, or do not 
satisfy the CLT.
The case $p=2$ and $q=1,2$ corresponds to the total energy and residual 
variance estimates for residual sampling  given in the previous two sections.

This analysis is limited to expectations that can be expressed in terms of the 
local energy field variable.
It is possible to generalise the analysis given to estimates of other 
quantities in VMC, since expectation values of operators are generally 
formulated as expectations of field variables (the local energy in the 
previous analysis) over the physical PDF of the system (the $\lambda \psi^2$ in 
the above).
This can always be reformulated through a change of random variables to provide 
the estimate as a mean of a lower dimensional PDF.

\section{Other estimates}
\label{sec:otherestimates}
It has been shown \cite{trail07a} that for standard sampling the CLT fails and 
the generalised central limit theorem takes its place for a variety of 
estimates of physical quantities.
This is a direct consequence of singularities appearing in the sampled field 
variable, and may be dealt with using alternative sampling.

An ideal estimator would be one for which the strongest form of the CLT 
provides confidence intervals for the estimated quantity.
Two complementary approaches to creating such estimators naturally suggest 
themselves.
A first method (essentially that described in the preceding sections for total 
energy and residual variance estimates) is to choose a new sampling strategy 
such that power law tails in the sampled quantities are removed.
A second method is to construct an alternative estimator by adding terms to the 
sampled quantity that have a mean of zero, hence preserving the large sample 
size limit of the estimated quantity, but modifying the distribution of random 
error that occurs for finite sample size.
Both these approaches play a role in controlling the statistical error for 
general estimates.

One of the most basic physical quantities for which accurate estimates are 
required is the kinetic energy of a system (the electronic kinetic energy for 
the examples considered here).
Estimates of this are straightforward to construct in terms of a MC estimate of 
integrals.
Unfortunately, the integrand generally possesses singularities on 
hyper-surfaces in $3N$-dimensional space and so uncontrolled random errors 
occur in the form of power law tails in PDFs.

The most direct kinetic energy estimate is provided by the operator in the 
Hamiltonian, and takes the form
\begin{equation}
\mathsf{A}_r \left[ E_{KE} \right]
   =  \frac{\sum_{n=1}^r w(\mathsf{E}_n) \mathsf{K}_n}
           {\sum_{n=1}^r w(\mathsf{E}_n)                                   },
\label{eq:15}
\end{equation}
where
$\mathsf{K}_n = \left[ -\frac{1}{2} \psi^{-1} \nabla^2_{\mathbf{R}}\psi 
\right]_{\text{\sffamily\bfseries R}_n}$ 
is a local kinetic energy at a random sample point, 
${\text{\sffamily\bfseries R}_n}$, in $3N$-dimensional space, and $w=1$ 
corresponds to standard sampling.
This local kinetic energy possesses singularities for an electron approaching a 
nucleus, for an electron approaching another electron, and at the nodal 
surface, referred as type 1, 2, and 3 in \cite{trail07a} (this is true for any 
$\psi$ for which the Kato cusp conditions are satisfied, and for which 
$\nabla^2_{\mathbf{R}}\psi \neq 0$ on the nodal surface).
For standard sampling, these singularities remain present in the sampled 
quantity, and the CLT is weakly valid in the sense that $x^{-4}$ asymptotic 
tails are present in the PDF of the estimate for finite sample size. 
For residual sampling, type 3 singularities are removed, but types 1 and 2 
remain, hence again the CLT is weakly valid.
In both cases the error is dominated by the presence of singularities of 
types 1 and 2, and these are unavoidable in the sense that they will be present 
for the exact wavefunction.

Green's $1^{st}$ theorem provides the means to remove the type 1 and 2 
singularities, giving  a new estimate of the form
\begin{equation}
\mathsf{A}_r \left[ E_{KE} \right]
   =  \frac{1}{2}
      \frac{\sum_{n=1}^r w(\mathsf{E}_n) \mathsf{F}^2_n}
           {\sum_{n=1}^r w(\mathsf{E}_n)                          },
\label{eq:16}
\end{equation}
where
$\mathsf{F}^2_n = \frac{1}{2}\left[ \sum_i \mathbf{F}_i.\mathbf{F}_i 
\right]_{\text{\sffamily\bfseries R}_n}$,
 with the sum over all electrons, and $\mathbf{F}_i = \psi^{-1} \nabla_i \psi $ 
the drift velocity vector of  electron $i$.

\begin{figure}[t]
\includegraphics{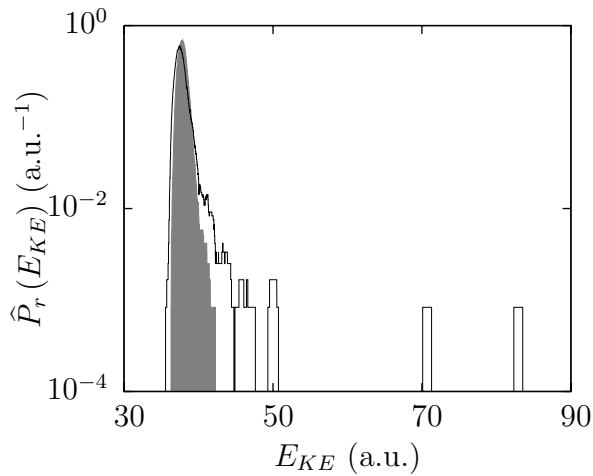}
\caption{\label{fig9}
Estimated PDFs for kinetic energy estimates constructed from different sampling 
strategies.
The unfilled solid curve is for standard sampling, and the grey filled curve 
for residual sampling.
In both cases, a kernel estimate of the PDF was constructed from $10^3$ 
kinetic energy estimates, with each kinetic energy estimate constructed 
from $r=10^3$ samples.
}
\end{figure}

The distribution of the random error in the estimate for both the standard and 
residual sampling case can be obtained in the same way as for the total energy 
and residual variance estimates considered previously.
The sole difference is in the order of the singularities present in the 
averaged field variable.
For standard sampling the analysis shows that the sum of random variables that 
make up the estimate does not obey the CLT, and an infinite variance Stable 
distribution with $x^{-5/2}$ tails results.
For residual sampling, the summed random variables are bounded, hence all 
co-moments exist, the bivariate CLT is valid in its strongest form, and 
Fieller's theorem provides a confidence interval for the estimated kinetic 
energy.
Figure~\ref{fig9} shows a kernel estimate of the PDF for kinetic energy 
estimates of the same carbon trial wavefunction described earlier.
The figure explicitly shows the failure of the CLT for standard sampling, and 
the improved estimate resulting from residual variance sampling, both using 
Eq.~(\ref{eq:16}).

If we compare the results from the two different types of estimator, 
Eq.~(\ref{eq:16}) with residual sampling provides $E_{KE}=37.894(17)$ a.u., 
whereas Eq.~(\ref{eq:15}) with standard sampling gives $E_{KE}=37.879(48)$ a.u.
Standard sampling requires eight times as many samples as residual sampling to 
provide the same accuracy for kinetic energy estimates, and, in addition, to 
obtain the confidence intervals for standard sampling it must be assumed that 
enough samples have been taken for the power law tails to be unimportant.

Finally, we note that residual sampling can only handle singularities at the 
nodal surface.
For many estimates a `transfer' of singularities with types 1 and 2 to the 
nodal surface may be achieved using `zero-variance, zero-bias' corrections of 
the form described by Assaraf and Caffarel\cite{assaraf99,assaraf03}.
However, there may be quantities for which estimates that possess no type 1 or 
type 2 singularities are unavailable.
Estimates of such quantities may be still be constructed using a more general 
sampling strategy defined by the estimator
\begin{equation}
\mathsf{A}_r \left[ 
    \frac{\langle \psi | \hat{G} | \psi\rangle}{\langle \psi | \psi \rangle} 
               \right]
   =  \frac{1}{2}
      \frac{\sum_{n=1}^r w(\mathsf{G}_n) \mathsf{G}_n  }
           {\sum_{n=1}^r w(\mathsf{G}_n)               },
\end{equation}
where 
$\mathsf{G}_n = \left[ \psi^{-1} \hat{G} \psi 
\right]_{\text{\sffamily\bfseries R}_n}$.
The sampling strategy would be defined by choosing $w$ to be a function of 
$\mathsf{G}_n$ that ensures that the summands are bounded, all co-moments 
exist, and so the strongest limit theorems apply.

\section{Conclusion}
\label{sec:conc}
 Previously it has been shown that the distribution of statistical errors in 
the estimates of the two most important basic quantities of variational QMC, 
provided by the most common `standard sampling' implementation of the method, 
result in an uncontrolled statistical error.
This results in the presence of unexpected outliers in estimates, and the 
failure of the CLT.
Here a more general sampling strategy is used, referred to as 
`residual sampling'.
Residual sampling prevents the artificial introduction of singularities in the 
sampled quantities that is an inherent part of the standard sampling strategy, 
and the accompanying statistical difficulties.
The new sampling strategy reintroduces the CLT for the total energy and 
residual variance in a strong form such that the deviation of the distribution 
from Normal for finite sample size is known and is bounded.

The `cost' of residual sampling is that the local energy must be evaluated in 
order to generate sample points with the required distribution, increasing 
computational expense, and that the interpretation of the random error in 
estimates is more complicated as the estimate must be considered as a quotient 
of two correlated random variables, rather than a single random variable.

The price of computational cost and complexity may be justifiable for 
estimating the total energy.
Numerical results for an isolated all-electron carbon atom suggest that 
residual sampling provides a modest improvement in the error of the estimated 
total energy for the all-electron carbon atom considered, since for this case 
leptokurtotic power law tails are weak for achievable sample sizes.
However, it should be borne in mind that these tails may be stronger for other 
systems, cannot be accurately (that is without bias) estimated, and are 
completely removed by residual sampling.

The increases in cost and complexity is justifiable for estimating the residual 
variance, since residual sampling provides a qualitative as well as 
quantitative improvement to estimates.
The analysis and numerical data clearly shows that residual sampling provides a 
controlled and small random error, unlike the standard sampling case.
This approach to controlling the random error in estimates is also expected to 
be important for other physical quantities - the CLT has been shown to be 
invalid for several estimates \cite{trail07a} and residual sampling, or a 
variant of residual sampling, provides a natural approach to achieving a Normal 
distribution of random error.

A primary application of the sampling strategies described is expected to be 
the optimisation of trial wavefunctions.
A considerably smaller number of samples are expected to be required to obtain 
an accurate minimum, since the random error of the optimised quantity is not 
Normal for standard sampling but is described by a bivariate Normal 
distribution for residual sampling.
Residual sampling also does not require the introduction of \textit{ad hoc} 
stabilisation methods, such as weight limiting\cite{kent99}.
A further feature of the new sampling strategy is that it samples the trial 
wavefunction close to the nodal surface - the standard sampling method avoids 
sampling here - the region where the accuracy of the trial wavefunction 
influences the accuracy of subsequent Diffusion Monte Carlo (DMC) 
calculations\cite{foulkes01}.

An analysis of the statistical errors of estimated quantities in VMC has not 
previously been available in the literature.
An assumption of a valid CLT has repeatedly been relied upon to justify methods 
and results, for both the estimation of physical quantities and optimisation of 
trial wavefunctions.
The analysis and residual sampling approach described here provide a method for 
predicting the random errors in QMC, and designing new sampling strategies that 
control and reduce the random error.
It also provides the possibility of preferentially optimising a trial 
wavefunction in the region of the nodal surface, and so providing a new means 
to control the fixed node error of DMC methods.

\begin{acknowledgements}
The author thanks Prof. Richard Needs for helpful discussions, and
financial support was provided by the Engineering and Physical
Sciences Research Council (EPSRC), UK.
\end{acknowledgements}


\end{document}